# Identifying Potential Risks and Benefits of Using Cloud in Distributed Software Development

Nilay Oza[1], Jürgen Münch[1]

Juan Garbajosa[2], Agustin Yague[2], Eloy Gonzalez Ortega[3]

[1]University of Helsinki – Finland, [2]Universidad Politecnica Madrid – Spain, [3]Indra Software Labs - Spain

`{nilay.oza,juergen.muench}@cs.helsinki.fi, egonza-`
`lezort@indra.es, {jgs, agustin.yague@upm.es}`

**Abstract.** Cloud-based infrastructure has been increasingly adopted by the industry in distributed software development (DSD) environments. Its proponents claim that its several benefits include reduced cost, increased speed and greater productivity in software development. Empirical evaluations, however, are in the nascent stage of examining both the benefits and the risks of cloud-based infrastructure. The objective of this paper is to identify potential benefits and risks of using cloud in a DSD project conducted by teams based in Helsinki and Madrid. A cross-case qualitative analysis is performed based on focus groups conducted at the Helsinki and Madrid sites. Participants' observations are used to supplement the analysis. The results of the analysis indicated that the main benefits of using cloud are rapid development, continuous integration, cost savings, code sharing, and faster ramp-up. The key risks determined by the project are dependencies, unavailability of access to the cloud, code commitment and integration, technical debt, and additional support costs. The results revealed that if such environments are not planned and set up carefully, the benefits of using cloud in DSD projects might be overshadowed by the risks associated with it.

## 1  Introduction

Cloud computing is no longer a new phenomenon. It is now ubiquitous in the industry and is increasingly used to develop software in a distributed environment. In many cases, it seems an obvious choice, especially for new, emerging ecosystems where companies want to involve external developers or external development teams in their own development projects or programs. However, whether cloud-based distributed software development (DSD) really works to address a wide range of technical and non-technical issues in DSD settings has not yet been validated scientifically. In other words, the expected benefits and risks of developing software in the cloud with distributed teams must be better understood.

Based on the literature [2, 12], we define cloud as the delivery of a stack of hardware or software residing in the data centre as a utility-like service over the network. We particularly focus on DSD in the context of cloud-based computing environments. We refer to DSD in the cloud as "software that is developed on a cloud-based platform across geographically distributed sites in a multi-stakeholder ecosystem."

To examine cloud-based DSD, we conducted an industry-led development project across three multi-site DSD teams. In this paper, we focus solely on what worked (i.e., was beneficial) with respect to cloud and the consequent risks attached to its use in the project. In the DSD project, we present here, our scope is limited to private cloud. Additional analyses of this project are planned or have been performed in parallel, such as the challenges involved in adding a new team to an on-going cloud-based distributed project. The findings of these other analyses are beyond the scope of this paper and will be published separately.

Section 2 describes work related to this paper. Section 3 presents a qualitative case study, including the research questions, the research approach, the context of the study, and the case company. Section 4 presents the results of the analysis of the case data. Section 5 stipulates the limitations and concerns about the validity of the study. Finally, Section 6 summarizes the study and provides an outlook on future research.

## 2    Related Work

DSD is often carried out in an ecosystem comprised of developers, clients, users, and other key stakeholders. Any new technology adoption will significantly affect such an ecosystem when several stakeholders are involved [17]. The cloud is not different in the sense that it would set DSD at new level of development within a complex multi-stakeholder, people-centric ecosystem [2, 6, 16, and 17].

Technically, cloud enables elasticity, scalability, and flexibility in DSD teams, resulting in overall increased productivity. For example, cloud-based DSD allows teams to perform rapid testing, dynamically scale up or down the required computing infrastructure, and produce working software updates rapidly [13]. Financially, cloud enables cost savings, faster time to market, and benefits of scale [7]. These claims, however, need further empirical validation, particularly because DSD-specific challenges may remain and even intensify in cloud-based development across distributed teams. The results of a few previous studies have already shown the benefits and risks of using cloud in DSD-specific environments [e.g., 2, 4, 8 and 17]. For example, Hashmi et al. [8] used cloud to facilitate DSD challenges, claiming that it would result in benefits for the infrastructure, platform, and provision of software as a service. They [8] also raised several concerns about using cloud in DSD, such as determining different levels of needs in service provision at distributed locations, the availability and subscription of cloud-based services for different types of dependency relationships among cloud users (tenants), and conducting project knowledge transfers across DSD sites. Phaphoom et al. [13] examined a practitioner forum on the benefits of using cloud in software development, finding that practitioners seemed to focus more effort on understanding how to utilize the dynamic scaling of cloud-based resources. Based on the prototypes showcased for using cloud in DSD, Yara et al. [17], claimed that although the hype about cloud has caused some fear, uncertainty, and doubt (FUD), this infrastructure will bring significant benefits to all key stakeholders in the DSD ecosystem. They also acknowledged other concerns, such as vendor lock-in, SLA control, privacy, reliability, data migration and access, auditing, and norms of regulation compliance.

Further empirical studies conducted in projects using cloud in DSD are required to establish systematically the merits of using cloud in this environment [8, 13 and 17]. In this paper, we contribute to the understanding of using cloud in DSD settings by studying the benefits and risks experienced by two distributed development teams.

## 3      Case Study

In this section, we present the overall setting of the qualitative study and research methodology adopted.

### 3.1     Research Question

We present our findings based on our answer to the following research question:

RQ – What are the potential benefits (what works well) and risks (what does not work well) of using a cloud-based infrastructure in a DSD project?

The answer to RQ will focus on the potential benefits and risks of using a cloud-based platform in a DSD project.

### 3.2     Context

The software factory (i.e., experimental research setting) is a novel software engineering research and education laboratory at the University of Helsinki. It offers a unique setting to conduct applied empirical investigations. All projects in the Software Factory originate from the industry's needs, and their duration is seven weeks. The software factory concept has been adopted by several other universities and companies. The software factories at the University of Helsinki, the Technical University of Madrid (UPM), and Indra are the development sites of the distributed project presented in this paper.

A qualitative study was performed in the context of a distributed development project that included three sites: 1) a software factory at the Indra Software Labs in Madrid, Spain; 2) a software factory at the Technical University of Madrid (UPM), Spain; and 3) a software factory at the University of Helsinki, Finland. The study reports the investigation of the Indra-led DSD project, which used a cloud-based platform. The Helsinki-based team was added for a seven-week development cycle. The qualitative interviews and supplementary data from the Helsinki team (also referred to as the "new team" in the project) are the main focus of the analysis in the proposed study. Working together, the Spanish and Helsinki teams developed solutions for the required massive data analysis by using Hadoop. MapReduce was used for large data calculations. A relevant issue is that the product owner required the quality profile of the resulting product to be commercially usable. Otherwise, the trial would not have been useful. It is also noteworthy that the product owner in the project was from Indra Software Labs. Although this requirement imposed strong pressure on the development teams, it served to reinforce the validity of the results. The project included five project members from the Indra site, seven project members from the UPM site, and eight project members from the Helsinki site. The project also had two dedicated agile coaches, one at Helsinki and one at UPM. The project included one product owner from Indra and three researchers actively involved during the collaborative period. Every engineer that participated in the software factory project had at least two years of software development experience. Based on interviews, they were recruited to align their competencies with the project's needs. An iterative development using the Kanban-based software development process was used [10]. During the seven-week project cycle, seven weekly sprints were conducted, and customer demos and retrospectives were provided after each sprint.

### 3.3 Research approach

We use Runeson and Höst's (2009) [15] guidelines for conducting case studies, which comprise five major steps, including study design, preparation for data collection, collecting evidence, analysis of the collected data, and reporting. We also considered validity concerns during the case study.

Regarding the design of the study, the qualitative nature of the examination justified the exploratory nature of the inquiry. The study used the distributed teams across three sites as the units of analysis. The unit of observation was the period of seven weeks, when all teams worked together in the project. Using semi-structured interviews, the direct data were collected from two focus groups, the teams from the Helsinki and Spanish sites. In the semi-structured interviews, topics were provided and approximate times were allocated for each topic. An interview guide was developed to assist the researcher during the interviews. A tape-recorded notice was given. The researcher also ensured the confidentiality and anonymity of the collected information as per the project's research agreement. All the interviews were audio recorded and later transcribed. The researchers also took notes during the interviews when they deemed something particularly relevant. Krueger and Casey's [11] guidelines were followed in the focus group sessions. The indirect data, which consisted of participant observations and notes from retrospective sessions, were stored as a text narrative received from the observer. They were used as complementary sources of information during the data analysis.

We analyzed the transcribed interviews using the "editing approach," recommended by Runeson and Höst [15] for software engineering case studies. In the editing approach, codes are defined based on findings of the researcher during the analysis. A preliminary set of codes was derived and applied to the transcripts. The codebook was then developed. Each statement in the transcribed interviews was given a unique identification and classified by two researchers. The transcribed data was then entered in tables, allowing for the analysis of patterns in the data. The encoding was done using an open coding method [11]. Each statement was analyzed and linked to the codebook. The percentages of statements and linked codes were calculated to generate visualizations of the themes.

## 4 Results

In this section, we present the key findings of our analysis. We focus mainly on specific findings related to DSD in the cloud infrastructure rather than generic DSD-related issues. In addition, we classify and present the answers to the research question according to the information revealed in the data set.

### 4.1 RQ: What are the key benefits and risks of using a cloud-based infrastructure in a DSD project?

Our study revealed five major benefits of using a cloud-based platform: rapid development, continuous integration, cost savings, code sharing, and faster ramp-up. In addition, it revealed five major risks of using a cloud-base infrastructure: dependencies, unavailability of access to the cloud, code commitment and integration, technical debt, and additional support costs. Fig. 2 and Fig. 3 provide overviews of the number of times the risks and benefits were mentioned in the qualitative interview.

#### 4.1.1 Key benefits of using the cloud in DSD

**Rapid development.** One of the major benefits experienced by the teams in the project was sharing cloud-based tools across distributed teams. As soon as the team obtained access to the cloud, it was ready to contribute to the development. One of the participants stated "As we did not have to replicate the technical environment, we were able to start development right after the required access was available to the remote infrastructure."

**Continuous integration.** The centrality of software development in the cloud brought with it the benefit of continuous integration. Software development in the cloud has a unique benefit in the sense that developers can commit frequent deliveries, even if the other team or unit is not in control of the development environment. One participant commented: "We were directly accessing the remote server where the product code integration happened. We just committed all codes to the main repository."

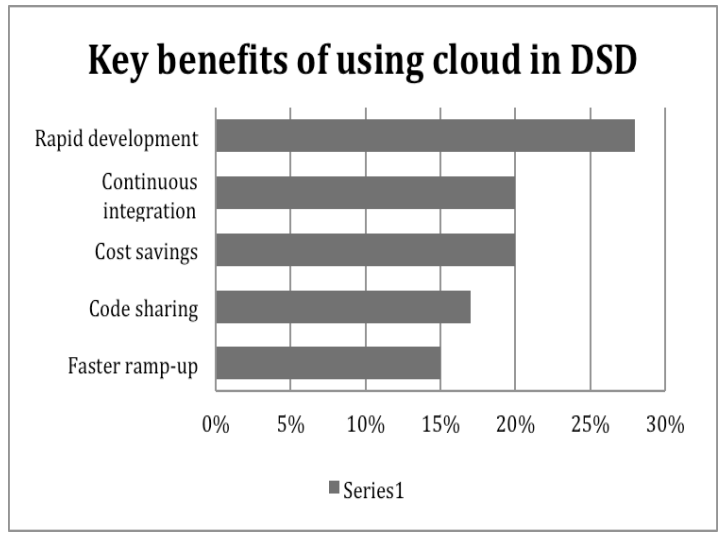

**Fig. 1.** Key benefits of using cloud in DSD

**Cost savings.** Because of near-instant access to development resources and no need to install and configure several tools on local machines, substantial costs of hardware and software were saved at multiple sites.

**Code sharing.** The cloud-based platform facilitated the sharing of codes across teams. Although the teams did not use concurrent distributed programming tools, the Spanish team shared the codebase with the Helsinki team, which had the required access controls.

**Faster ramp-up.** The team felt that although several issues of coordination and organization were encountered in the initial stages of the project, the progress occurred quite rapidly after the initial period. Once the technical environment was understood, the access to cloud-based resources proved very useful in speeding up the software development. One participant said "It took two weeks to really settle on a shared understanding, but then work started, and they (Spanish team) started to see that we could contribute quite a lot…We started to get more work then."

### 4.1.2   Key risks of using the cloud in DSD

**Dependencies.** Dependencies, in terms of both technical and operational issues, created several challenges for the teams to work together. Our analysis indicated dependencies on at least two levels: operational and technical. For example, at times, one team had to wait for the other to catch up, provide feedback, or take specific actions before the development could move to the next stage. One participant said "We had to depend quite a lot on the Spanish team to get lot of things done—sometimes we went into waiting mode or were just not able to implement things, as we had to get some-thing done by the other team." The Spanish team member had a contradicting comment, however, "I saw them more like – 'give me work, I do my work, just my own work'. They were depending on us for one output, one use case." One reason behind this dependency was indicated in the resource imbalance between the Spanish and Helsinki teams, as described above. The Helsinki team worked full time for a seven-week period, whereas the Spanish team worked part time for three or four hours a day.

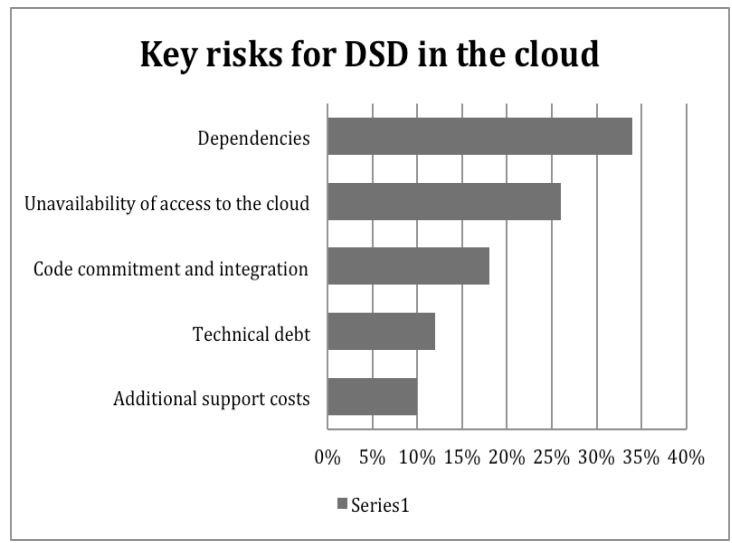

**Fig. 2.** Key risks of using cloud in DSD

One participant commented "We simply had more people with a full-time work commitment at our end. This meant we could do much more than the other teams thought we could. Probably, some information mismatch happened on how much work capacity was available at our end." Our results also revealed technical dependencies on the codebase as well as the overall complexity of the product's technical environment. This was reflected in the following comment: "There were many applications in the central codebase with linkages to components, etc. It was difficult for us to determine the dependencies of these inter-connected modules, mainly because of the lack of interaction between teams."

**Unavailability of access to the cloud.** The new team, which did not have direct control of the cloud-based platform, were challenged by the lack of accessibility to necessary resources. The results indicated that the new team had to rely completely on the Spanish team to gain access to the cloud-specific resources if they were not previously allocated. The general challenge that we observed was also reflected in some of the team members' comments at the Helsinki site: "The remote server is not controlled by us. If and when we lose access to it (for whatever reason), we have to contact the

Spanish team, and only they can re-establish access for us." Another comment revealed a similar challenge: "It is quite difficult to continue working when the team loses access to the cloud. Because of the common codebase and central integration, we have to wait until we are able to get access." The teams also experienced increase in uncertainties when the common cloud-based software was inaccessible. One of the Spanish members commented "infrastructures, when fail, they generate uncertainty. Also, they have created dependency; within these limits, there was a certain dependency. For example, one day, Redmine crashed – and it becomes difficult to keep track on the user histories with the acceptation criteria."

**Code commitment and integration.** The study showed that committing code to the proper code repository could be challenging, particularly if the cloud-based platform was not fully known to the team. Similarly, integrating code with the overall product required additional testing on the cloud platform.

**Technical debt.** The study revealed that as multiple teams started to commit changes to the cloud-based platform, the consequent changes in other linked parts of the codebase were not visible. More specifically, it was difficult for the new team to see the evolving impact of the changes and additions to the codebase on the cloud-based platform. One of the participants said "We did not have to worry about the platform, as it was shared by all teams, but because of frequent releases and our lack of understanding of the overall product (at least initially), we had to leave certain changes undone although we thought they would be worth implementing." He further commented "We got specific errors in the build, and we could see that there were errors, but when we told them, they said the errors did not occur at their end".

**Additional support costs.** The study also showed that a cloud-based platform in DSD requires additional managerial and operational support. Although the cloud-based platform has several benefits, additional overhead should be considered.

## 5    Discussion and Limitations

The results of our study confirmed some earlier findings and provided new insights about using cloud in DSD. The benefits experienced by the distributed teams in our study are similar to those reported in Hasmi et al. [8] and Yara et al. [17]. In particular, cloud-enabling continuous integration in DSD supports Yara et al.'s [17] hypothesis that the cloud could play a "game-changing role" in software testing. The benefits identified in our study also specifically increase the understanding of how cloud could fit into DSD settings, which are highly people-centric, multi-stakeholder ecosystems.

The results specific to the potential risks of using cloud in DSD provided new insights. In particular, potential risks related to dependencies of the DSD teams in addition to the potential increase in technical debt in the software were specifically identified in our study.

Our study extended our initial inquiry into using cloud in a DSD setting as well as if and to what extent it is beneficial. Although the latter questions are beyond the scope of the present study, we could uncover empirically founded insights into the merits of using cloud in DSD.

Our study brings novelty in terms of empirically founded approach to studying benefits and risks of using cloud in DSD and also demonstrating that adoption of cloud still may pose similar threats that were also experienced in DSD projects not based on cloud. One could argue that our findings are potentially the same as they are usually experienced in DSD setting. We, however, note that our study aims to empirically identify benefits and risks of using cloud in DSD rather than claiming any distinct benefits or risks of cloud vs. DSD.

### 5.1 Limitations

We considered the four validity concerns recommended by [15] in conducting case studies in software engineering. We also followed suggestions for improving the validity of our study [15]: triangulation, developing and maintaining a detailed case study protocol; review of designs, protocols, etc. by peer researchers; review of collected data and obtained results by the case subjects; and spending sufficient time on the case.

One of the limitations of our study is related to subjectivity in the collected qualitative data, which calls into question the validity of our findings. Using the focus group and detailed case study protocol, however, increased confidence in the qualitative data because they were gathered from several persons. In terms of external validity, another limitation concerns the validity of our findings, which resides in the weak generalizability of the results. Although the results could be of high interest to those involved in cloud-based DSD outside the investigated case, the generalizability of the present findings requires further, similar empirical studies. However, the detailed case study protocol and analysis by multiple researchers increased the reliability and replicability of our study in other cloud-based DSD settings, which would help strengthen the generalizable conclusions regarding potential benefits and risks of using cloud in DSD. Furthermore, triangulation was achieved in multiple ways—data were collected from both direct and indirect sources, multiple researchers were used to develop a reliable coding scheme, and case representatives reviewed the results. Although the study was of limited duration, this potential limitation was reduced by the fact that the researchers had a long-term connection with the organization before the present case study was implemented.

## 6 Conclusions and Future Work

This paper presented a qualitative study that aimed to identify the key benefits and risks of using a cloud-based platform in a DSD project to better understand if and how cloud works in a DSD setting. The study findings indicated that several of the expected merits of cloud computing could be utilized in real DSD projects. In particular, the findings indicated increased development speed, easier integration of development activities, and simplified access to development resources. However, the qualitative analysis also revealed that inherent challenges of DSD, such as lack of informal communication, temporal differences, need for work synchronization, and lack of trust, persist and must be addressed by special measures beyond cloud. A combination of cloud and other measures is needed to address these challenges successfully. In a wider perspective, the use of cloud in DSD would clearly have beneficial effects on development processes and even lead to new types of development processes. These effects are based on advanced cloud's enhanced possibilities for developing software, providing software to customers, and obtaining feedback from customers (e.g., possibilities for continuous integration, continuous global deployment, and live customer feedback). We plan to replicate the study and do further analyses of the data to examine DSD in the cloud to determine what works well and what does not.

### 6.1   Acknowledgments

We sincerely thank our industry partner, Indra Software Lab, for setting up and running the presented DSD project with research-centric interests. We also thank the Finnish technology agency, Tekes, for funding the Cloud Software Factory project, the Cloud Software Program, and the SCABO project, under which the proposed research study was undertaken. Finally, we thank the Spanish Ministry of Science and Innovation for partially sponsoring the project IPT-430000-2010-38 i-Smart Software Factory under the INNPACTO Program.

### 6.2   References

1. Arimura, Y., Ito, M. Cloud Computing for Software Development Environment. ―In-house Deployment at Numazu Software Development Cloud Center― .Fujitsu Sci. Tech. J, 47(3)325-334. (2011).
2. Armbrust, M., Fox, A., Griffith, R., Joseph, A. D., Katz, R., Konwinski, A., Lee, G., Patterson, D., Rabkin, A., Stoica, I. and Zaharia, M. A view of cloud computing. Communications of the ACM, 53 (4), 50–58. (2010).
3. Buyya, R., Yeo, S., Venugopal, S., Broberg, J., and Brandic, I. Cloud computing and emerging IT platforms: Vision, hype, and reality for delivering computing as the 5th utility, Future Generation Computer Systems. 25(6). 599-616. (June 2009). ISSN 0167-739X, 10.1016/j.future.2008.12.001.
4. Dillon, T., Wu, C., and Chang, E. Cloud Computing: Issues and Challenges. 24th IEEE International Conference on Advanced Information Networking and Applications. (2010)
5. Erdogmus, H.: Cloud Computing: Does Nirvana Hide behind the Nebula? IEEE Software. 26(2). 4-6. (March-April 2009) doi: 10.1109/MS.2009.31
6. Garrison, G., Kim, S., and Wakefield. R.L.: Success factors for deploying cloud computing. Communications of ACM. 55, 9. (September 2012) 62-68
7. Grossman, R.L. The Case for Cloud Computing. IT Professional. 11(2). 23-27. (March-April 2009). doi: 10.1109/MITP.2009.40
8. Hashmi, S.I.; Clerc, V.; Razavian, M.; Manteli, C.; Tamburri, D.A.; Lago, P.; Nitto, E.D.; Richardson, I.: Using the Cloud to Facilitate Global Software Development Challenges.

Global Software Engineering Workshop (ICGSEW) - 2011 Sixth IEEE International Conference on. (Aug 2011) 70-77. doi: 10.1109/ICGSE-W.2011.19'
9. Kim, W., Kim, D.S., Lee, E., Lee, S.: Adoption issues for cloud computing. In: Proceedings of iiWAS2009, pp. 3-6 (Dec 2009).
10. Kniberg, H.: Kanban and Scrum - Making the Most of Both. Lulu.com. 2010
11. Krueger, R.A, Casey, M.A.: Focus groups: a practical guide for applied research. Pine Forge Press. (2009)
12. Mell, P., and Grance, T. Draft - NIST working definition of cloud computing - v15. (Aug 2009). [URL: http://www.nist.gov/itl/cloud/upload/cloud-def-v15.pdf]
13. Phaphoom, N., Oza, N., Wang, X., and Abrahamsson. P. Does cloud computing deliver the promised benefits for IT industry?. In *Proceedings of the WICSA/ECSA 2012 Companion Volume* (WICSA/ECSA '12). ACM, New York, NY, USA, 45-52. (2012). DOI=10.1145/2361999.2362007
14. Rimal, B.P., Choi, E., Lumb, I.: A Taxonomy and Survey of Cloud Computing Systems. NCM '09. Fifth International Joint Conference on. 44-51. (Aug. 2009). doi: 10.1109/NCM.2009.218
15. Runeson P., and Höst, M.: Guidelines for conducting and reporting case study research in software engineering. Empirical Software Engineering. 14(2). (2009). 131-164. DOI: 10.1007/s10664-008-9102-8
16. Voas, Jeffrey; Zhang, Jia; , "Cloud Computing: New Wine or Just a New Bottle?," *IT Professional*. 11(2). 15-17. (March-April 2009). doi: 10.1109/MITP.2009.23
17. Yara, P., Ramachandran, R., Balasubramanian, G., Muthuswamy, K., Chandrasekar, D. Global Software Development with Cloud Platforms. Software Engineering Approaches for Offshore and Outsourced Development Lecture Notes in Business Information Processing. In O. Gotel, M. Joseph, and B. Meyer (Eds.): SEAFOOD 2009, LNBIP 35, 81–95. Springer-Verlag Berlin Heidelberg (2009)